\documentclass[a4paper,11pt]{article}
  
\usepackage{physics}
\usepackage{appendix}

\usepackage[top=2.cm, left=2.cm, right=2.cm, bottom=2.cm]{geometry}

\usepackage{amsfonts,amssymb,amsmath,amsthm,dsfont}
\usepackage{xfrac}
\usepackage{tensor}

\usepackage{subfigure}
\usepackage{xcolor}
\usepackage{tikz} 
\usetikzlibrary{external}
\makeatletter
\let\pgfmathModX=\pgfmathMod@
\let\pgfmathdivX=\pgfmathdiv@
\usepackage{pgfplots}%

\let\pgfmathMod@=\pgfmathModX
\let\pgfmathdiv@=\pgfmathdivX
\makeatother

\usepackage{pgfplotstable}
\usepgfplotslibrary{fillbetween}

\usepackage{graphicx}
\usepackage{verbatim} 

\makeatletter
\let\NAT@parse\undefined
\makeatother
\usepackage[sort&compress, numbers]{natbib}

\usepackage{hyperref}

\usepackage[footnotesize]{caption}

\newcommand{\ie}{\emph{i.e.}, }

\newcommand{\etal}{\emph{et al. }}

\newcommand{\Christoffel}[4]{\frac{1}{2} g^{#1 #4}\left(g_{ #4 #2,#3} + g_{#4 #3 ,#2} - g_{#2 #3,#4} \right)}

\newcommand{\ChristoffelWeylianPotentialSimplify}[5]{\bar \Gamma^{#1}_{#2 #3} + \frac{1}{2}  \left(\ln(#5)_{,#3} \delta^{#1}_{#2} + \ln(#5)_{,#2}\delta^{#1}_{ #3} - \ln(#5)_{,#4} \bar g^{#1#4} \bar g_{#2 #3} \right)}

\newcommand{\ConstFrontR }{\gamma \hbar^2}

\title{\vspace{-1cm} Gravity as a purely quantum effect \\[-3mm]}
\author{Beno\^{i}t Pairet$^1$$^\dagger$\\
\footnotesize $^1$ISPGroup/ICTEAM,
UCLouvain, Belgium,\\ [-4mm]
}
\date{\empty}

\renewenvironment{abstract}{\noindent\bf\small {\em Abstract---}}{}

\begin{document}

\maketitle

\begin{abstract}
General relativity and quantum mechanics are perhaps the two most successful theories of the XX$^{\text{th}}$ century. Despite their impressive accurate predictions, they are both valid at their own scales and do not seem to be expressible using the same framework. It is commonly accepted that in order to create a consistent theory of both quantum mechanics and gravity, it is required to quantize the gravitational field.

In the present paper, another path is taken on which the Einstein field equations emerge from a geometric formulation of relativistic quantum mechanics. In this context, there appears to be no need for quantizing gravity since gravity would in fact be a fully quantum effect.

\let\thefootnote\relax\phantom{\footnotemark}\footnotetext{
\scriptsize  $^*$ BP is funded by the Belgian FNRS, under a FRIA grant (Boursier FRIA). Part of this study is funded by the AlterSense project (MIS-FNRS).}
\vspace{-1mm}
\end{abstract}

\section{Introduction}

Quantum gravity designates a research program that aims to find a theory that unifies quantum mechanics and gravity. It is widely accepted that quantizing gravity is a requirement for such a theory. However, quantizing gravity proves to be a most challenging task, and as of today, there exists no fully consistent theory of quantum gravity. For a recent analysis of the difficulties encountered when attempting to quantize gravity, as well as a historical presentation, see for instance~\cite{rovelli2000notes,carlip2015quantum}.

The modern theory of gravity is fully classical. It is encoded by the Hilbert-Einstein action
\begin{equation}
S = \int \left [\frac{1}{2\kappa} R + \mathcal L_M \right ] \sqrt{-g}d^4x.
\label{HE_action}
\end{equation}
When~\eqref{HE_action} is varied with respect to $g^{\mu \nu}$, one recovers the Einstein field equations
\begin{equation}
R_{\mu \nu} + \frac{1}{2} R g_{\mu \nu} =  \kappa T_{\mu \nu}.
\label{EFE}
\end{equation}
In~\eqref{HE_action} and~\eqref{EFE}, $R_{\mu \nu}$ is the Ricci tensor, $R$ is the scalar curvature,  $\kappa = 8 \pi G/c^4$ is a constant, $\mathcal L_M$ is the matter part of the Lagrangian, and $T_{\mu \nu} = -2\frac{\delta \mathcal L_M}{\delta g^{\mu\nu}} + g_{\mu\nu} \mathcal L_M$ is the stress-energy tensor (see for instance~\cite{khriplovich2005general}).

It is a common practice to construct a quantum theory by quantizing a classical theory. For instance, quantizing the norm of the 4-momentum
\begin{equation}
p_\mu p^{\mu}  = m^2c^2,
\end{equation}
leads to the Klein-Gordon equation. This is done by recasting $p_\mu$ as the operator $\hat P_{\mu} =i\hbar \partial_\mu$ acting on $\Psi$. One then obtains
\begin{equation}
\partial_\mu \partial^{\mu} \Psi + \frac{m^2c^2}{\hbar^2} \Psi = 0,
\label{regualarKGequation}
\end{equation}
which is indeed the Klein-Gordon equation, describing a scalar field (see for instance~\cite{greiner1990relativistic}). Hence if one wishes to build a theory of quantum gravity, it seems that the natural way to do so is to quantize the Einstein field equations. 

Except if gravity is already a quantum theory. Clearly, the last sentence may appear as an absurdity. However, work by Santamato~\cite{santamato1984geometric,santamato1984statistical,santamato1985gauge} show that quantum effects can be considered as purely geometric. Quantum mechanics is described as classical mechanics in a non-Euclidean geometry. The quantum effects appear as a subtle interplay between the particles and the structure of spacetime. We hence recover the idea from John Archibald Wheeler that \textit{``spacetime tells matter how to move; matter tells spacetime how to curve"}. 

With this in mind, the assertion that gravity may be a quantum effect is no longer absurd. The aim of this paper is precisely to show that the Einstein field equations, in fact, emerge from the Lagrangian of geometric quantum mechanics. 

The rest of the paper is as follows. In Section~\ref{sec:SantamatoGeometric}, we briefly introduce Santamato's geometric version of the Klein-Gordon equation. We also take some time to show that the spacetime remains Riemannian. In Section~\ref{sec:derivationGR}, Santamato's action is slightly modified to make it well defined on the whole spacetime. Then, the Einstein field equations are derived from this modified action. In Section~\ref{sec:discussion}, we discuss some implications of the present theory. The microscopic and macroscopic limits are found. Finally, we argue that a clear ontology for quantum effects is required. We conclude in Section~\ref{sec:conclusion} where we mention some perspectives and future work.
 
\section{Santamato's geometric Klein-Gordon equation}
\label{sec:SantamatoGeometric}

In a series of papers, Santamato presented a geometric formulation of quantum mechanics, first for the Schr\"odinger equation~\cite{santamato1984geometric} then for the Klein-Gordon equation~\cite{santamato1984statistical,santamato1985gauge}. More recently, Santamato and De Martini extended this geometric approach of quantum mechanics for 1/2 spin particles~\cite{santamato2013derivation}. The present work is solely concerned with the geometric Klein-Gordon equation. The case of spin particles is left for future work.

In Santamato's formulation, quantum mechanics emerges as classical mechanics in a Weylian spacetime~\cite{weyl1918gravitation}. However, as it is discussed in Section~\ref{Rgauge}, we already stress that the underlying of spacetime remains Riemannian. In the present section, we briefly summarize Santamato's theory of spinless relativistic quantum mechanics~\cite{santamato1984statistical,santamato1985gauge}. For a concise introduction of Weylian geometry, see for instance Section~2 of~\cite{rosen1982weyl}.

The basic idea of Weylian geometry is that when a vector is parallel transported, not only its orientation but also its length is allowed to change. Under a parallel transport, the length of a vector $\xi$ is changed by
\begin{equation}
d \xi = \xi \phi_\mu dx^\mu,
\end{equation} 
where $\phi_\mu$ is a geometrical field called Weylian potential. The Weylian affine connexion is given by
\begin{equation}
\Gamma^\lambda_{\mu\nu} = \{^\lambda_{\mu\nu} \} +\frac{1}{2} (\phi_\mu \delta ^\lambda_\nu +\phi_\nu \delta ^\lambda_\mu - g_{\mu\nu}g^{\lambda\rho}\phi_\rho ),
\label{WeylianConnexion}
\end{equation}
where $\{^\lambda_{\mu\nu} \}$ are the Christoffel symbols. Another important feature of Weylian geometry is the invariance under a conformal gauge transform. This aspect is discussed in Section~\ref{Rgauge} where it allows one to show that the geometry of spacetime remains Riemannian. 

In~\cite{santamato1985gauge}, the following Lagrangian for a spinless particle is presented (assuming there is no external field, \ie the particle is free)
\begin{equation}
L(x,dx) = (m^2c^2 - \ConstFrontR R)^{1/2} ds,
\label{lagrangian_line}
\end{equation}
where $m$ is the mass of the particle, $\gamma = 1/6$ and $R$ is the spacetime scalar curvature. From this results a family of hypersurfaces $S(x)=\text{constant}$, where $S(x)$ obeys the Hamilton-Jacobi equation
\begin{equation}
g^{\mu \nu} \partial_{\mu} S \partial_{\nu} S = m^2c^2 - \ConstFrontR  R.
\label{eq:HJE_S_R}
\end{equation}
A congruence of curves intersect the family of hypersurfaces $S$,
\begin{equation}
\frac{dx^{\mu} }{ ds } = \frac{g^{\mu \nu} \partial_{\nu} S}{(g^{\rho \sigma} \partial_{\rho} S \partial_{\sigma}S)^{1/2}}.
\label{eq:KG_ParticleTraj}
\end{equation}
A probability current density $j^\mu$ is associated with the congruence of curves, and is given by
\begin{equation}
j^\mu = \rho \sqrt{-g}g^{\mu\nu} \partial_\nu S,
\label{eq:KG_Current}
\end{equation}
with some $\rho>0$. The current $j^\mu$ is a conserved quantity
\begin{equation}
\partial_\mu j^\mu = 0.
\label{eq:KG_divLessCurrent}
\end{equation}

Santamato obtains the spacetime affine connection from the Lagrangian~\eqref{lagrangian_line} by varying the action with respect to the fields $\Gamma^{\lambda}_{\mu\nu}$. In order to do so, the action is put into the four-volume integral
\begin{equation}
\int_{\Omega} [(m^2c^2 - \ConstFrontR  R)(g_{\sigma \alpha}j^\sigma j^\alpha)]^{1/2} d^4x,
\label{lagragian_4D_around_Particle}
\end{equation}
where $\Omega$ is the spacetime region occupied by the particle. The affine connexion is found to be the Weylian connexion~\eqref{WeylianConnexion} with the Weylian potential $\phi_\mu = \partial_\mu \ln(\rho)$. Hence, the positive scalar $\rho$ is a geometrical field.

The equivalence between Santamato's geometric quantum mechanics (GQM) and the Klein-Gordon equations is shown in~\cite{santamato1985gauge}. We briefly present it here for the sake of completeness. First notice that inserting $\Psi = \sqrt{\rho} \exp{-iS/\hbar}$ in Equation~\eqref{regualarKGequation} and separating real and imaginary parts, the Klein-Gordon equation becomes
\begin{align}
&\frac{\sqrt{\rho}}{\hbar} \left(\frac{\partial_\mu \rho \partial^\mu S}{\rho} + \partial_\mu \partial^\mu S \right) = 0,\label{KG_equation_rho_S__1} \\
&  \partial_\mu S \partial^\mu S = m^2c^2 + \frac{\hbar^2}{2} \left( \frac{\partial_{\mu}\partial^{\mu} \rho}{\rho} -\frac{1}{2}\frac{\partial_\mu \rho\partial^\mu \rho}{\rho^2}  \right) .\label{KG_equation_rho_S__2}
\end{align}

Second, we show that equations of Santamato's GQM lead to Equations~\eqref{KG_equation_rho_S__1} and~\eqref{KG_equation_rho_S__2}. The scalar curvature $R$ is given by~\cite{weyl1922space}
\begin{equation}
R = R_m - 3 \left(\frac{1}{2} g^{\mu \nu} \phi_\mu \phi_\nu + (1/\sqrt{g}) \partial_\mu (\sqrt{-g} g^{\mu \nu}\phi_\nu)\right),
\label{eqScalarCurv}
\end{equation}
where $R_m$ is the Riemannian scalar curvature, \ie the scalar curvature constructed from the Christoffel symbols. Considering a reference frame where $g_{\mu\nu}$ is the Minkowski metric $\eta_{\mu \nu}$,~\eqref{eqScalarCurv} reads
\begin{align}
R  & = - 3 \left(   \frac{\partial^\mu  \partial_\mu \rho}{\rho} -  \frac{1}{2} \frac{\partial^\mu \rho \partial_\mu \rho}{\rho^2 }\right).
\label{eqScalarCurv_minkowski_rho}
\end{align}
Inserting~\eqref{eqScalarCurv_minkowski_rho} into~\eqref{eq:HJE_S_R} yields the second Klein-Gordon equation~\eqref{KG_equation_rho_S__2}. Furthermore with $g_{\mu\nu} = \eta_{\mu \nu}$,~\eqref{eq:KG_divLessCurrent} and~\eqref{eq:KG_Current} is just the first Klein-Gordon equation~\eqref{KG_equation_rho_S__1}.

\subsection{Riemannian gauge}
\label{Rgauge}
An important property of Weyl geometry is that it is invariant under the conformal change
\begin{equation}
g_{\alpha\beta} \rightarrow g'_{\alpha\beta}  = e^{\Lambda(x)}g_{\alpha\beta},
\label{gaugeMetric}
\end{equation}
as long as the Weyl potential is simultaneously changed to
\begin{equation}
\phi_{\mu} \rightarrow \phi_{\mu} - \Lambda(x)_{,\mu}.
\label{gaugeWeylPotential}
\end{equation}
The simultaneous transforms~\eqref{gaugeMetric} and~\eqref{gaugeWeylPotential} are referred to as gauge transform. 

Since for GQM, the Weyl potential is given by a total derivate, $\phi_\mu = \partial_\mu \ln(\rho)$, we can impose $\Lambda(x)_{,\mu}=  \partial_\mu \ln(\rho)$ so that the Weylian potential is zero everywhere. According to~\eqref{gaugeMetric}, the metric becomes
\begin{equation*}
g_{\alpha\beta}  = e^{\ln(\rho) + k} \bar g_{\alpha\beta} = \rho e^{k} \bar g_{\alpha\beta}.
\end{equation*}
Since in this gauge, the Weylian potential is zero everywhere, the connexion is then fully determined by the Christoffel symbols and the geometry is Riemannian. Hence, we refer to the choice $\Lambda(x)_{,\mu}=  \partial_\mu \ln(\rho)$ as the Riemannian gauge. 

We note that if $\rho=0$, then the metric vanishes. This implies that $\rho>0$ everywhere. In the Riemannian gauge, the scalar field $\rho$ is considered as a conformal factor defined on the whole spacetime. 

It is important to stress that indeed the geometry is left unchanged in the Riemannian gauge. That is readily seen by computing the Christoffel symbols with $g_{\alpha\beta} =\rho e^{k} \bar g_{\alpha\beta}$. This yields
\begin{align*}
\Gamma_{\sigma \nu}^{\rho} &=\Christoffel{\rho}{\sigma}{\nu}{\alpha}\\
& = \frac{1}{2} \frac{1}{\rho} \bar g^{\rho \alpha}\left( (\rho \bar g_{ \alpha \sigma})_{,\nu} + (\rho \bar g_{\alpha \nu})_{ ,\sigma} - (\rho \bar g_{\sigma \nu})_{,\alpha} \right) \\
& = \ChristoffelWeylianPotentialSimplify{\rho}{\sigma}{\nu}{\alpha}{\rho},
\end{align*}
where $x_{,\nu} = \partial_{\nu} x$ and $\bar \Gamma^{\rho}_{\sigma\nu}$ are the Christoffel symbols constructed with $\bar g_{\alpha\beta}$. One can see that the Riemannian connexion is the same as~\eqref{WeylianConnexion}. 

The connexion can be thought as having two components. First the metrical component  $\bar \Gamma^{\rho}_{\sigma\nu}$ and then a scalar field component  $\frac{1}{2}\left(\ln(\rho)_{,\nu} \delta^{\rho}_{\sigma} + \ln(\rho)_{,\sigma}\delta^{\rho}_{ \nu} - \ln(\rho)_{,\beta} \bar g^{\rho\beta} \bar g_{\sigma \nu} \right)$. Since the scalar field component depends on the conformal factor $\rho$, this component is also referred to as the conformal component.

\section{Derivation of the Einstein field equations}
\label{sec:derivationGR}

In the present work, we modify the action \eqref{lagragian_4D_around_Particle} into
\begin{equation}
I = \int [(m^2c^2 f(x) - \ConstFrontR R)(g_{\sigma \alpha}j^\sigma j^\alpha)]^{1/2} d^4x,
\label{lagragian_4D}
\end{equation}
where $f(x)$ is a function that encodes the presence/absence of particles. The precise form of $f(x)$ is intentionally left unspecified and will be briefly discussed in Section~\ref{sec:conclusion}.
 The introduction of $f(x)$ allows the action \eqref{lagragian_4D} to be integrated on all spacetime and not only on $\Omega$.

With the action~\eqref{lagragian_4D}, Equation~\eqref{eq:HJE_S_R} becomes
\begin{equation}
g^{\mu \nu} \partial_{\mu} S \partial_{\nu} S = m^2c^2 f(x) - \ConstFrontR  R.
\label{eq:HJE_S_mf_R}
\end{equation}

As it is done with the Hilbert-Einstein action, the action \eqref{lagragian_4D} is now varied with respect to $g^{\mu \nu}$.

\begin{align*}
\frac{\delta}{\delta g^{\mu \nu}} I & = \frac{\delta}{\delta g^{\mu \nu}}  \int [(m^2c^2 f(x) - \ConstFrontR R)(g_{\sigma \alpha}j^\sigma j^\alpha)]^{1/2} d^4x\\
& =  \int \frac{\delta}{\delta g^{\mu \nu}}[(m^2c^2 f(x) - \ConstFrontR R)(g_{\sigma \alpha}j^\sigma j^\alpha)]  \frac{1}{[(m^2c^2 f(x) - \ConstFrontR R)(g_{\sigma \alpha}j^\sigma j^\alpha)]^{1/2}}d^4x\
\end{align*}

We set $A = \frac{1}{[(m^2c^2 f(x) - \ConstFrontR R)(g_{\sigma \alpha}j^\sigma j^\alpha)]^{1/2}}$ and we continue
\begin{align*}
\frac{\delta}{\delta g^{\mu \nu}} I & = \int  \frac{\delta}{\delta g^{\mu \nu}}[(m^2c^2 f(x) - \ConstFrontR R)(g_{\sigma \alpha}j^\sigma j^\alpha)]  Ad^4x\\
& = \int  \left (
(m^2c^2 \frac{\delta}{\delta g^{\mu \nu}}f(x) - \ConstFrontR  \frac{\delta}{\delta g^{\mu \nu}}R)(j^\alpha j_\alpha)
+(m^2c^2 f(x) - \ConstFrontR R)\frac{\delta}{\delta g^{\mu \nu}}(j^\alpha j_\alpha)  \right )
Ad^4x\\
& = \int  \left (
(m^2c^2 \frac{\delta}{\delta g^{\mu \nu}}f(x) - \ConstFrontR R_{\mu \nu})(j^\alpha j_\alpha)
+(m^2c^2 f(x) - \ConstFrontR R)\frac{\delta}{\delta g^{\mu \nu}}(j^\alpha j_\alpha)  \right)
Ad^4x
\end{align*}

Inserting the value of $j^\alpha j_\alpha$ from Equations~\eqref{eq:KG_Current} and~\eqref{eq:HJE_S_mf_R}, \ie \[ j^\alpha j_\alpha = \rho^2 \sqrt{-g}^2 \partial^\alpha S \partial_\alpha S =  \rho^2 \sqrt{-g}^2 (m^2c^2 f(x) - \ConstFrontR R ),\] yields
 \begin{align*}
\textstyle  \frac{\delta}{\delta g^{\mu \nu}} I  = 
& \textstyle\int  \left ( \left (m^2c^2 \frac{\delta}{\delta g^{\mu \nu}}f(x) - \ConstFrontR R_{\mu \nu} \right) \left(\rho^2 \sqrt{-g}^2 (m^2c^2 f(x) - \ConstFrontR R )\right) +(m^2c^2 f(x) - \ConstFrontR R)\frac{\delta}{\delta g^{\mu \nu}}(j^\alpha j_\alpha)  \right) Ad^4x\\
=& \textstyle \int  \left(\rho^2 \sqrt{-g}^2\left(m^2c^2 \frac{\delta}{\delta g^{\mu \nu}}f(x) - \ConstFrontR R_{\mu \nu}\right) +\frac{\delta}{\delta g^{\mu \nu}}(j^\alpha j_\alpha) \right) 
 \left(m^2c^2 f(x) - \ConstFrontR R \right)Ad^4x\\
 =& \textstyle \int  
\left(\rho^2 \sqrt{-g}^2\left(m^2c^2 \frac{\delta}{\delta g^{\mu \nu}}f(x) - \ConstFrontR R_{\mu \nu} \right)
+\frac{\delta}{\delta g^{\mu \nu}}(j^\alpha j_\alpha) \right) 
\frac{ m^2c^2 f(x) - \ConstFrontR R }{[(m^2c^2 f(x) - \ConstFrontR R)(g_{\sigma \alpha}j^\sigma j^\alpha)]^{1/2}}d^4x\\
=& \textstyle  \int  
\left(\rho^2 \sqrt{-g}^2\left(m^2c^2 \frac{\delta}{\delta g^{\mu \nu}}f(x) - \ConstFrontR R_{\mu \nu}\right)
+\frac{\delta}{\delta g^{\mu \nu}}(j^\alpha j_\alpha) \right) 
\frac{ m^2c^2 f(x) - \ConstFrontR R }{[(\rho^2 \sqrt{-g}^2 (m^2c^2 f(x) - \ConstFrontR R)^2 ]^{1/2}}d^4x\\
= & \textstyle\int  
\left(\rho^2 \sqrt{-g}^2\left(m^2c^2 \frac{\delta}{\delta g^{\mu \nu}}f(x) - \ConstFrontR R_{\mu \nu}\right)
+\frac{\delta}{\delta g^{\mu \nu}}(j^\alpha j_\alpha) \right) 
\frac{ m^2c^2 f(x) - \ConstFrontR R }{\rho \sqrt{-g} (m^2c^2 f(x) - \ConstFrontR R)}d^4x\\
= & \textstyle\int  
\left(\rho^2 \sqrt{-g}^2\left(m^2c^2 \frac{\delta}{\delta g^{\mu \nu}}f(x) - \ConstFrontR R_{\mu \nu}\right)
+\frac{\delta}{\delta g^{\mu \nu}} \left (\rho^2 \sqrt{-g}^2 (m^2c^2 f(x) - \ConstFrontR R ) \right ) \right) 
\frac{ 1 }{\rho \sqrt{-g} }d^4x.
\end{align*}

The term $B = \frac{\delta}{\delta g^{\mu \nu}} \left (\rho^2 \sqrt{-g}^2 (m^2c^2 f(x) - \ConstFrontR R ) \right ) $ is given by
\begin{align*}
\textstyle B & \textstyle=
\left( \frac{\delta}{\delta g^{\mu \nu}}(\rho^2) \sqrt{-g}^2
+\rho^2 \frac{\delta}{\delta g^{\mu \nu}}(\sqrt{-g}^2) \right) (m^2c^2 f(x) - \ConstFrontR R )
+\rho^2 \sqrt{-g}^2 \frac{\delta}{\delta g^{\mu \nu}}(m^2c^2 f(x) - \ConstFrontR R )\\
& \textstyle=
\rho^2 \frac{\delta}{\delta g^{\mu \nu}}(-g) (m^2c^2 f(x) - \ConstFrontR R )
+\rho^2 \sqrt{-g}^2 (m^2c^2 \frac{\delta}{\delta g^{\mu \nu}}f(x) - \ConstFrontR \frac{\delta}{\delta g^{\mu \nu}}R ),
\end{align*}
where we used $\frac{\delta}{\delta g^{\mu \nu}}\rho^2 = 0$.

We have that $\delta g = g g^{\mu \nu} \delta g_{\mu \nu}$ and that $  g^{\mu \nu} \delta g_{\mu \nu} = - g_{\mu \nu} \delta g^{\mu \nu} $. Combining the two yields $\frac{\delta}{\delta g^{\mu \nu}}(-g) = -(-g) g_{\mu \nu}$. Inserting this into $B$ yields
\begin{align*}
\textstyle B& \textstyle=
-\rho^2 (-g) g_{\mu \nu} \left(m^2c^2 f(x) - \ConstFrontR R \right)
+\rho^2 \sqrt{-g}^2  \frac{\delta}{\delta g^{\mu \nu}} \left(m^2c^2 f(x) - \ConstFrontR R \right) \\
&  \textstyle= \rho^2 (-g) \left(- g_{\mu \nu} \left(m^2c^2 f(x) - \ConstFrontR R \right)
+ m^2c^2 \frac{\delta}{\delta g^{\mu \nu}}f(x) - \ConstFrontR  R_{\mu \nu} \right ).
\end{align*}
Inserting the value of $B$ into the variation yields
\begin{align*}
 \textstyle \frac{\delta}{\delta g^{\mu \nu}} & \textstyle= \int  
\rho \left(m^2c^2 \frac{\delta}{\delta g^{\mu \nu}}f(x) - \ConstFrontR R_{\mu \nu}
- g_{\mu \nu} \left(m^2c^2 f(x) - \ConstFrontR R \right)
+ m^2c^2 \frac{\delta}{\delta g^{\mu \nu}}f(x) - \ConstFrontR  R_{\mu \nu}  \right)  \sqrt{-g} d^4x \\
& \textstyle = \int  
\rho \left( 2 m^2c^2 \frac{\delta}{\delta g^{\mu \nu}}f(x) - 2 \ConstFrontR R_{\mu \nu}
-  g_{\mu \nu} m^2c^2 f(x) + g_{\mu \nu}  \ConstFrontR R 
  \right)  \sqrt{-g} d^4x.
\end{align*}
Requiring that the action is stationary with respect to variations of $g^{\mu \nu}$ leads to
\begin{align}
& \textstyle \frac{\delta}{\delta g^{\mu \nu}} I  = 0 \nonumber \\ 
\textstyle \Leftrightarrow \quad & \textstyle \int  
\rho \left( 2 m^2c^2 \frac{\delta}{\delta g^{\mu \nu}}f(x) - 2 \ConstFrontR R_{\mu \nu}
-  g_{\mu \nu} m^2c^2 f(x) + g_{\mu \nu}  \ConstFrontR R 
  \right)  \sqrt{-g} d^4x = 0 \nonumber \\
\textstyle\Leftrightarrow \quad \textstyle & 2 m^2c^2 \frac{\delta}{\delta g^{\mu \nu}}f(x) - 2 \ConstFrontR R_{\mu \nu}
-  g_{\mu \nu} m^2c^2 f(x) + g_{\mu \nu}  \ConstFrontR R 
= 0 \nonumber \\
\textstyle \Leftrightarrow \quad &     \textstyle R_{\mu \nu} -  g_{\mu \nu}  \frac{1}{2}R = \left(2\ConstFrontR \right)^{-1} \left( 2 m^2c^2 \frac{\delta}{\delta g^{\mu \nu}}f(x) 
-  g_{\mu \nu} m^2c^2 f(x) \right) \nonumber \\
\textstyle \Leftrightarrow \quad &   \textstyle  R_{\mu \nu} -  g_{\mu \nu}  \frac{1}{2}R =  2 \gamma \bar \kappa \bar T_{\mu \nu},
\label{QuantumEFE}
\end{align}
where we set $\bar \kappa =\frac{c^2}{\hbar^2}$ and $\bar T_{\mu \nu} = \left( 2 m^2c^2 \frac{\delta}{\delta g^{\mu \nu}}f(x) 
-  g_{\mu \nu} m^2c^2 f(x) \right)$.

\subsection{Units and constants}

Equation~\eqref{QuantumEFE} resembles the Einstein field equations. We note that both $\bar \kappa$ and $\bar T_{\mu \nu}$ do not have the units of $\bar \kappa$ and $T_{\mu \nu}$ respectively. The stress-energy tensor $T_{\mu \nu}$ has units of MT$^{-2}$L$^{-1}$ and the constant $\kappa$ has units of L$^{-1}$M$^{-1}$T$^2$. We readily check that Equations~\eqref{HE_action} and~\eqref{EFE} are consistent since $\kappa T_{\mu \nu}$ has units of L$^{-2}$, which are the units of both the Riemannian tensor and the curvature scalar.

Furthermore, we notice that for \eqref{lagragian_4D} and \eqref{eq:HJE_S_mf_R} to be consistent, $f(x)$ must be unitless. This implies that $\bar T_{\mu \nu}$ has units of M$^2$ and it is thus not a stress-energy tensor. However, its definition is similar to the definition of the stress-energy tensor. We wish to make this similarity clearer. To that end, we use the Planck units to change the units of $\bar T_{\mu \nu}$ into the ones of $T_{\mu \nu}$.

We multiply $\bar T_{\mu \nu}$ by $m_{\text{P}}^{-1}\cdot t_{\text{P}}^{-2}\cdot l_{\text{P}}^{-1} (8\pi)^{-1}$ where $m_{\text{P}}$, $t_{\text{P}}$, and $l_{\text{P}}$ are respectively the Planck mass, the Planck time, and the Planck length. They are given by~\cite{mohr2016codata}
\begin{equation}
 m_{\text{P}} = \sqrt{\frac{\hbar c}{G}}, \quad \quad
 t_{\text{P}} = \sqrt{\frac{\hbar G}{c^5}},\quad \quad
 l_{\text{P}}= \sqrt{\frac{\hbar G}{c^3}}.
\end{equation}
In order for Equation~\eqref{QuantumEFE} to remain consistent, we multiply $\bar \kappa$ by $8\pi m_{\text{P}}\cdot t_{\text{P}}^{2}\cdot l_{\text{P}} $. Then $\bar \kappa$ becomes 
\begin{align}
\bar \kappa &= 8\pi\sqrt{\frac{\hbar c}{G}} \left(\sqrt{\frac{\hbar G}{c^5}}\right)^2 \sqrt{\frac{\hbar G}{c^3}} \frac{c^2}{\hbar^2},\\
&= 8\pi \frac{G}{c^4} = \kappa.
\end{align}

Equation~\eqref{QuantumEFE} now reads $R_{\mu \nu} -  g_{\mu \nu}  \frac{1}{2}R =  \kappa \cdot 2 \cdot \gamma \cdot m_{\text{P}}^{-1}\cdot t_{\text{P}}^{-2}\cdot l_{\text{P}}^{-1}  (8\pi)^{-1} \bar T_{\mu \nu}$. Setting $T_{\mu \nu} = 2 \cdot \gamma \cdot m_{\text{P}}^{-1}\cdot t_{\text{P}}^{-2}\cdot l_{\text{P}}^{-1} (8\pi)^{-1} \bar T_{\mu \nu}$, we recover the Einstein field equations. 

We can summarize the result of the present section as
\begin{equation}
\frac{\delta}{\delta g^{\mu \nu}} \int [(m^2c^2 f(x) - \ConstFrontR R)(g_{\sigma \alpha}j^\sigma j^\alpha)]^{1/2} d^4x =0 \Rightarrow R_{\mu \nu} -    \frac{1}{2}R g_{\mu \nu} = \kappa T_{\mu \nu}.
\end{equation}
This is a striking and rather astonishing result as we recovered the Einstein field equations from the action of relativistic GQM. This indicates that gravity and quantum effects would in fact be one and the same. Quantum mechanics could be considered as gravity at the microscopic level. Or one could argue that gravity can be thought as a macroscopic manifestation of quantum effects.

\section{Discussion}
\label{sec:discussion}
We now consider some important aspects of the present theory. First, the microscopic and macroscopic limits are considered. Second, we argue that an epistemological analysis of the physical world is required. Indeed, in a physical theory that includes both quantum mechanics and gravity, concepts such as trajectories must have a precise meaning. The Heisenberg uncertainty principle is also briefly discussed.

\subsection{Microscopic and macroscopic limits}

The minimal requirement of a new physical theory is that experimentally well established theories are recovered as limiting cases. With this in mind, one expects to recover both regular quantum mechanics and general relativity as the microscopic and macroscopic limits of the following equations 
\begin{align}
& g^{\mu \nu} \partial_{\mu} S \partial_{\nu} S = m^2c^2 f(x) - \ConstFrontR  R,\label{mM_HamiltonJacobi}\\
& \partial_{\mu} \left (\rho \sqrt{-g}g^{\mu\nu} \partial_\nu S \right) = 0,\label{mM_Cons_current}\\
& R_{\mu \nu} -   \frac{1}{2}R g_{\mu \nu} = \left(2\ConstFrontR \right)^{-1} \left( 2 m^2c^2 \frac{\delta}{\delta g^{\mu \nu}}f(x) 
-  g_{\mu \nu} m^2c^2 f(x) \right)\label{mM_EFE}.
\end{align}

Concerning the microscopic limit, we argue that most quantum experiments are performed in local reference frames of small laboratories on Earth. Because of the equivalence principle, it is always possible to choose coordinates such that the metric is locally flat. 

This implies that for all practical purposes of the quantum experiment, one can consider that the metric is the Minkowski metric. We already showed that in this case,~\eqref{mM_HamiltonJacobi} and~\eqref{mM_Cons_current} yield the Kelin-Gordon equation. 

As for~\eqref{mM_EFE}, we show that it trivially holds with a Minkowski metric and thus we can ignore~\eqref{mM_EFE} all together. The argument goes as follows. First, observe that $\frac{\delta f}{\delta \eta^{\mu \nu}}=0$. Second, injecting~\eqref{mM_HamiltonJacobi} into~\eqref{mM_EFE} with $g_{\mu \nu} = \eta_{\mu \nu}$ yields
\begin{align}
R_{\mu \nu} &= \frac{1}{2} R \eta_{\mu \nu} -  \frac{1}{2} (\ConstFrontR)^{-1} m^2c^2 f(x) \eta_{\mu \nu}\\
& = \frac{1}{2} (\ConstFrontR )^{-1} (m^2c^2f(x)  - \partial^{\sigma} S \partial_{\sigma} S)\eta_{\mu \nu} -  \frac{1}{2} (\ConstFrontR)^{-1} m^2c^2 f(x) \eta_{\mu \nu} \\
& =  - \frac{1}{2} (\ConstFrontR)^{-1} \partial^{\sigma} S \partial_{\sigma} S \eta_{\mu \nu}. 
\label{Temp_results_EFE}
\end{align}
Third, re-injecting~\eqref{Temp_results_EFE} onto~\eqref{mM_EFE} gives
\begin{align}
\partial^{\sigma} S \partial_{\sigma} S  \eta_{\mu \nu} =  (m^2c^2 f(x) - \ConstFrontR R  ) \eta_{\mu \nu},
\end{align}
which is simply~\eqref{mM_HamiltonJacobi} multiplied by $\eta_{\mu\nu}$. Hence, in flat space,~\eqref{mM_EFE} is implied by~\eqref{mM_HamiltonJacobi} and we thus fully recover ``regular" spinless relativistic quantum mechanics.

In the macroscopic world, it is not possible to choose a reference frame such that the metric is flat everywhere. This implies that~\eqref{mM_EFE} holds, as expected at a macroscopic scale. Since the macroscopic limit is generally given by $\hbar \rightarrow 0$,~\eqref{mM_HamiltonJacobi} becomes
\[ g^{\mu \nu} \partial_{\mu} S \partial_{\nu} S = m^2c^2 f(x),\]
which is the classical Hamilton-Jacobi equation in a spacetime with metric $g_{\mu \nu}$. 

Hence, both microscopic and macroscopic limits are recovered from the present theory. However, there is an obvious physical difficulty that was not discussed. Indeed, quantum mechanics is usually considered as a random theory and, in particular, trajectories are not well defined. These aspects of quantum mechanics are known to be problematic when considering the macroscopic limit. Indeed macroscopic bodies possess well defined, deterministic trajectories. The aim of the next section is to shed some light on this apparent shortcoming. This is done mostly by linking the present theory to the Bohmian interpretation of quantum mechanics (see for instance~\cite{durr2009bohmian}). 

\subsection{Trajectories, uncertainty: a case for Bohmian mechanics}

In the Copenhagen interpretation of quantum mechanics, particles properties such as position and momentum do not have a precise meaning until a measurement is made. It is often argued that this is how the quantum world works and it does not in fact matter. After all the quantum world does not have to be explained by our classical intuition. As long as the classical and quantum worlds do not coexist within a theory, trying to make sense of quantum mechanics in terms of particle trajectories may indeed appear as unnecessary.

In the present theory however, there is no clear distinction between the classical and the quantum worlds. Since the Einstein field equations are recovered from quantum mechanics, the evolution of the universe can be accounted for by the same set of equations that describe the mechanics of particles (at least of spin 0). It then becomes clear that for such a theory to be consistent, the concept of quantum trajectories must be clarified. Also since it is not clear what the measurement process would be at the cosmological scale, one must take care of the quantum measurement problem.

Making sense of quantum mechanics is a task that is well beyond the scope of the present manuscript. Fortunately, there is no need to make sense of quantum mechanics since it has already been done. Indeed, Bohmian mechanics is a fully consistent theory that does not suffer from regular quantum paradoxes. All that needs to be done is to express the present theory in Bohmian terms and we then know that all the quantum effects can be consistently explained.

Since the present theory is relativistic, we will link it to relativistic Bohmian mechanics. However, Bohmian mechanics is manifestly non-local\footnote{Bohmian mechanics is in fact often rejected because of that. It is important to note that the non-locality of Bohmian mechanics is not in contradiction with Lorentz invariance. Furthermore, quantum mechanics \emph{is} non-local, as shown by the violation of Bell's inequalities~\cite{bell1964einstein}. And even if one considers quantum mechanics to be false, non-local entanglement effects have been experimentally observed~\cite{hensen2015loophole}. Hence, although physics laws are Lorentz invariant, they do exhibit some non-local phenomena.} and these non-local effects require some notion of simultaneity. This need of an expression of simultaneity is the motivation behind the Bohm-Dirac hypersurface model~\cite{durr1999hypersurface}. D\"urr \etal first introduce a foliation of spacetime by equal-time hypersurfaces $\Sigma$ such that the particles trajectories are perpendicular to these hypersurfaces. Then they argue that the instantaneous piloting of the particles by the wavefunction operates on these timelike hypersurfaces. In this settings, the particles motion is orthogonal to the hypersurface. The Bohm-Dirac hypersurface model is Lorentz invariant by construction. 

However, in~\cite{durr1999hypersurface}, there is no physical explanation for the source of the foliation. D\"urr \etal discuss a few possibilities but the question of the origin of the foliation remains unanswered. They merely point out that if one accepts such hypersurfaces, they allow to express Bohmian mechanics in a Lorentz compatible way. In this setting, the $k^{th}$ particle follows the trajectory
\begin{equation}
\frac{dX^{(k)}_\mu}{ds} = \frac{j^{(k)}_\mu}{j^{(k)}_\mu n^\mu},
\label{eq:HDBM_ParticleTraj}
\end{equation}
where $j^{(k)}_\mu$ is perpendicular to the hypersurface and $n^\mu$ is the unit vector normal vector of the hypersurface at the particule location.

To make clear the correspondence between Bohmian mechanics and GQM, we first point out that the scalar curvature given by~\eqref{eqScalarCurv_minkowski_rho} is the relativistic version of the quantum potential introduced by Bohm in his seminal paper~\cite{bohm1952suggested}. 

In Santamato's work, the current $j^{(k)}_\mu$ is by construction perpendicular to the hypersurface $S$. We can thus write $n_\mu$ as 
\[
\textstyle \frac{j^{(k)}_\mu }{  \sqrt{j^{(k)}_\sigma j^{(k)\sigma}} }.\] Then equation \eqref{eq:KG_ParticleTraj} becomes
\begin{equation}
\frac{dX^{(k)}_\mu}{ds}= \frac{j_k}{\sqrt{j^{(k)}_\sigma j^{(k)\sigma}}}.
\label{eq:HDBM_MOD_ParticleTraj}
\end{equation}
Now, injecting the current from $j_\mu$~\eqref{eq:KG_Current} onto equation~\eqref{eq:HDBM_MOD_ParticleTraj}, we get:
\begin{equation}
\textstyle \frac{dX^{(k)}_\mu}{ds} = \frac{\rho \sqrt{-g} \partial_\mu S}{\rho \sqrt{-g} \sqrt{  \partial_\sigma S\partial^\sigma S}} \bigg |_{(k)} 
= \frac{ \partial_\mu S}{ \sqrt{  \partial_\sigma S\partial^\sigma S}}\bigg |_{(k)},
\label{eq:HDBM_same_KG_ParticleTraj}
\end{equation}
where $|_{(k)}$ means that the expression is evaluated at the location of the $k^{th}$ particle. We see that~\eqref{eq:HDBM_same_KG_ParticleTraj} is the same expression as the congruence of curves~\eqref{eq:KG_ParticleTraj} evaluated at the location of the particle. Hence relativistic Bohmian mechanics and Santamato's GQM appear to be equivalent.

Concerning the uncertainty principle, Santamato gave it a geometric meaning in~\cite{santamato1988heisenberg}. A thorough analysis of Santamato's geometric incertitude principle is beyond the scope of the current work. However, it can be loosely stated as follows: the notion of length is modified by the presence of particles, hence it makes it impossible to measure positions with an infinite precision. 

The uncertainty principle is then a consequence of our inability to measure a particle state precisely. Bohmian mechanics offers the same view of the uncertainty principle. This reinforces the equivalence between Bohmian mechanics and Santamato's GQM and we can argue that the present theory has a well-defined macroscopic limit.

\section{Conclusion}
\label{sec:conclusion}

In this paper, we recovered the Einstein field equations from Santamato's geometric quantum mechanics. We derived the expected microscopic and macroscopic limits. We then pointed out that a clear ontology was required. We argued that linking the present theory to relativistic Bohmian mechanics provided the needed ontology.

In the present work, we solely considered spinless particles and extending the present theory to particles with spin would be an important next step. One way to do so would be to start with Santamato and De Martini work of 1/2 spin particles~\cite{santamato2013derivation}. Interestingly, they suggest that non-local effects are due to the interplay between the particles and the Weylian curvature scalar~\cite{de2014nonlocality}. Further investigations will most likely reveal interesting and perhaps unexpected properties of particles as well as their interactions with spacetime. 

Another very important extension of the present work would be the derivation of a geometric version of quantum field theory. In this derivation, the precise form of $f(x)$ should be made clearer. It could for instance be linked to the creation and annihilation operators. This is however well beyond the scope of the present paper.

And last but not least, it would be interesting to explore the implications -- if any -- of the present theory in cosmology. For instance the work~\cite{rosen1982weyl} and~\cite{deruelle2011conformal} showed that, when working in the Riemannian gauge, the universe can be considered as static. In that setting, the apparent inflation of the universe can be attributed to the fact that matter is continuously shrinking. It could be interesting to explore whether the observed acceleration of the expansion of the universe can be explained by this view. Indeed, one could argue that as time goes by, clusters of matter shrink, thus appearing to be further away. Hence, the attraction between two such clusters decreases with time, making the shrinking faster. This is however quite speculative and a thorough analysis of the cosmological implications of the present theory is required. 

\section*{Acknowledgements}

I would like to express my deepest gratitude to Prof. Laurent Jacques for For his very precious help and relentless support. 

I also would like to thank K\'evin Degraux, Simon Carbonnelle, St\'ephanie Gu\'erit, Pierre-Yves Gousenbourger, Valerio Cambareri, Fran\c{c}ois Rottenberg, Vincent Schellekens, Anne-Sophie Collin, Victor Joos, Antoine Vanderschueren, and Faustine Cantalloube for the many fruitful discussions we have had.

\small
\bibliography{biblio}{}

\begin{thebibliography}{10}

\bibitem{bell1964einstein}
John~S Bell.
\newblock On the einstein podolsky rosen paradox.
\newblock {\em Physics Physique Fizika}, 1(3):195, 1964.

\bibitem{bohm1952suggested}
David Bohm.
\newblock A suggested interpretation of the quantum theory in terms of" hidden"
  variables. i.
\newblock {\em Physical review}, 85(2):166, 1952.

\bibitem{carlip2015quantum}
Steven Carlip, Dah-Wei Chiou, Wei-Tou Ni, and Richard Woodard.
\newblock Quantum gravity: A brief history of ideas and some prospects.
\newblock {\em International Journal of Modern Physics D}, 24(11):1530028,
  2015.

\bibitem{de2014nonlocality}
Francesco De~Martini and Enrico Santamato.
\newblock Nonlocality, no-signalling, and bellʼs theorem investigated by weyl
  conformal differential geometry.
\newblock {\em Physica Scripta}, 2014(T163):014015, 2014.

\bibitem{deruelle2011conformal}
Nathalie Deruelle and Misao Sasaki.
\newblock Conformal equivalence in classical gravity: the example of "veiled"
  general relativity.
\newblock In {\em Cosmology, Quantum Vacuum and Zeta Functions}, pages
  247--260. Springer, 2011.

\bibitem{durr1999hypersurface}
Detlef D{\"u}rr, Sheldon Goldstein, Karin M{\"u}nch-Berndl, and Nino
  Zangh{\`\i}.
\newblock Hypersurface bohm-dirac models.
\newblock {\em Physical Review A}, 60(4):2729, 1999.

\bibitem{durr2009bohmian}
Detlef D{\"u}rr and Stefan Teufel.
\newblock {\em Bohmian mechanics: the physics and mathematics of quantum
  theory}.
\newblock Springer Science \& Business Media, 2009.

\bibitem{greiner1990relativistic}
Walter Greiner et~al.
\newblock {\em Relativistic quantum mechanics}, volume~3.
\newblock Springer, 1990.

\bibitem{hensen2015loophole}
Bas Hensen, Hannes Bernien, Ana{\"\i}s~E Dr{\'e}au, Andreas Reiserer, Norbert
  Kalb, Machiel~S Blok, Just Ruitenberg, Raymond~FL Vermeulen, Raymond~N
  Schouten, Carlos Abell{\'a}n, et~al.
\newblock Loophole-free bell inequality violation using electron spins
  separated by 1.3 kilometres.
\newblock {\em Nature}, 526(7575):682, 2015.

\bibitem{khriplovich2005general}
Iosif~Bent︠s︡ionovich Khriplovich.
\newblock {\em General Relativity}.
\newblock Springer Science \& Business Media, 2005.

\bibitem{mohr2016codata}
Peter~J Mohr, David~B Newell, and Barry~N Taylor.
\newblock Codata recommended values of the fundamental physical constants:
  2014.
\newblock {\em Journal of Physical and Chemical Reference Data}, 45(4):043102,
  2016.

\bibitem{rosen1982weyl}
Nathan Rosen.
\newblock Weyl's geometry and physics.
\newblock {\em Foundations of Physics}, 12(3):213--248, 1982.

\bibitem{rovelli2000notes}
Carlo Rovelli.
\newblock Notes for a brief history of quantum gravity.
\newblock {\em arXiv preprint gr-qc/0006061}, 2000.

\bibitem{santamato1984geometric}
Enrico Santamato.
\newblock Geometric derivation of the schr{\"o}dinger equation from classical
  mechanics in curved weyl spaces.
\newblock {\em Physical Review D}, 29(2):216, 1984.

\bibitem{santamato1984statistical}
Enrico Santamato.
\newblock Statistical interpretation of the klein--gordon equation in terms of
  the space-time weyl curvature.
\newblock {\em Journal of mathematical physics}, 25(8):2477--2480, 1984.

\bibitem{santamato1985gauge}
Enrico Santamato.
\newblock Gauge-invariant statistical mechanics and average action principle
  for the klein-gordon particle in geometric quantum mechanics.
\newblock {\em Physical Review D}, 32(10):2615, 1985.

\bibitem{santamato1988heisenberg}
Enrico Santamato.
\newblock Heisenberg uncertainty relations and average space curvature in
  geometric quantum mechanics.
\newblock {\em Physics Letters A}, 130(4-5):199--202, 1988.

\bibitem{santamato2013derivation}
Enrico Santamato and Francesco De~Martini.
\newblock Derivation of the dirac equation by conformal differential geometry.
\newblock {\em Foundations of Physics}, 43(5):631--641, 2013.

\bibitem{weyl1918gravitation}
Hermann Weyl.
\newblock Gravitation and electricity.
\newblock {\em Sitzungsber. K{\"o}nigl. Preuss. Akad. Wiss.}, 26:465--480,
  1918.

\bibitem{weyl1922space}
Hermann Weyl.
\newblock {\em Space--time--matter}.
\newblock Dutton, 1922.

\end{thebibliography}
\bibliographystyle{plain}

\end{document}